\documentclass[aps,prl,showpacs,twocolumn]{revtex4}
\usepackage{graphicx}
\input{epsf}

\begin{document}
\title {Enhancement of Josephson phase diffusion by microwaves}

\author{Y. Koval, M. V. Fistul\footnote{Present address:
Theoretische Physik III, Ruhr-Universit\"{a}t Bochum, D-44780
Bochum, Germany} and A. V. Ustinov}

\affiliation {Physikalisches Institut III, Universit\"at
Erlangen-N\"urnberg, D-91058, Erlangen, Germany}
\date{\today}
\begin{abstract}
We report an experimental and theoretical study of the phase
diffusion in small Josephson junctions under microwave
irradiation. A peculiar enhancement of the phase diffusion by
microwaves is observed. The enhancement manifests itself by a
pronounced current peak in the current-voltage characteristics.
The voltage position $V_{\rm top}$ of the peak increases with the
power $P$ of microwave radiation as $V_{\rm top}\propto\sqrt P$,
while its current amplitude weakly decreases with $P$. As the
microwave frequency increases, the peak feature evolves into
Shapiro steps with finite slope. Our theoretical analysis taking
into account the enhancement of incoherent superconducting current
by multi-photon absorption is in good agreement with experimental
data.
\end{abstract}


\maketitle

One of the most spectacular indications of the Josephson effect is
locking of Josephson oscillations to the frequency of external
microwaves. As the Josephson frequency is proportional to voltage,
the locking leads to appearance of steps at quantized voltages -
so called Shapiro steps \cite{Shapiro-63,Tinkham} - in the
current-voltage characteristic ($I$-$V$ curve) of a Josephson
junction. At the Shapiro step, the superconducting current flowing
through the junction oscillates \emph{coherently} with the
external microwave. In this letter, we describe a strong
\emph{incoherent} junction response to applied microwaves observed
in small mesoscopic Josephson tunnel junctions in the presence of
thermal fluctuations and dissipative environment.

Small mesoscopic Josephson junctions have received a great deal of
interest. They display such peculiar phenomena as the quantum and
thermal fluctuation-induced escape \cite{Tinkham,Clarke1}, Coulomb
blockade effects \cite{Aver-Likh}, Josephson phase diffusion
\cite{MartKautz,Ing-Grab-Eber-94}, to name just a few. The
significance of these various effects depends on the ratio of the
Josephson energy $E_{\rm J}$ to the charging energy $E_{\rm C}$
and on the thermal fluctuations. Here, we will deal with
properties of moderately small Josephson tunnel junctions, with
the ratio of $E_{\rm J}/E_{\rm C}\gg~1$, and $E_{\rm
J}\simeq~k_{\rm B} T $, which corresponds to the phase diffusion
regime induced by thermal fluctuations.

The Josephson phase diffusion in small junctions has been studied
in detail both experimentally \cite{MartKautz,Vion-96} and
theoretically \cite{IngNaz,GrIngPaul}. Recently, the phase
diffusion regime has been observed also in layered
high-temperature superconductors \cite{Hightemp}. The
characteristic feature of this regime is the absence of pure
zero-voltage superconducting state accompanied, however, by a
hysteresis in the $I$-$V$ curve. The phase diffusion behavior is a
consequence of frequency dependent damping \cite{MartKautz}, i.e.
the dissipation can be rather weak at zero frequency but it
reaches a substantial value at frequencies close to the plasma
frequency $\omega_p$ of the junction. That leads to a diffusion of
the junction state with time in the Josephson phase space.

The influence of microwave radiation on Josephson tunnel junctions
is, in general, well understood. There are such well-known effects
as Shapiro steps mentioned above, photon assisted tunneling of
quasiparticles \cite{Tinkham}, and resonant escape of the
Josephson phase \cite{Clarke1,FistWalUst}. However, the influence
of microwave radiation in the Josephson phase diffusion regime got
so far very little attention. While being interesting by itself,
the microwave-driven phase diffusion is also an implementation of
a very general phenomenon as the ac driven diffusion of a particle
in a periodic potential \cite{Hangi,Fistul,Vinokur}.

In this Letter, we report on novel behavior induced by microwaves
in small Josephson junctions. For low microwave frequencies, the
autonomous $I$-$V$ curves of the studied junctions display a
characteristic phase diffusion branch close to zero voltage, which
evolves into a pronounced peak shifting to higher voltages as the
power $P$ of the microwave radiation increases. We explain this
behavior as an enhancement of the Josephson phase diffusion by
microwaves. Our analysis carried out for relatively low microwave
frequencies $\omega~\ll~\omega_p$ takes into account both
incoherent superconducting current and multi-photon absorption
(emission). As the frequency of microwave radiation increases, a
crossover to well-separated Shapiro steps with a finite slope is
observed.

For the experiments, we have fabricated several sub-micron size
Nb/AlO$_x$/Nb Josephson tunnel junctions. The junction area was
defined in a sputtered Nb/AlO$_x$/Nb trilayer by using
electron-beam lithography and reactive ion etching. For insulation
of the trilayer edges we used highly cross-linked PMMA
\cite{Technology}. The critical current density of the trilayer is
about $220\,\rm{A/cm^2}$. We have measured several sub-micron
junctions of comparable dimensions, which all showed similar
behavior. Measurements presented in this paper have been performed
on a junction having the area of $0.07\,\rm{\mu m^2}$ as estimated
from its normal resistance $R_{\rm T}$. The Josephson coupling
energy calculated from the estimated fluctuation-free critical
current of $I_{\rm c}=150\,$nA (taken as a half of the current
rise at the gap voltage $V=2\Delta/e$, which is close to the
maximum expected Ambegaokar-Baratoff critical current $I_{\rm
AB}=\pi\Delta/(2eR_{\rm T})$) was $E_{\rm J}=I_{\rm
c}\Phi_0/(2\pi)=4.9\cdot 10^{-23}\,$J, where $\Phi_0=h/(2e)$ is
the magnetic flux quantum. We estimated the specific capacitance
$C_{\rm s}=46\,$fF/$\mu$m$^2$ of the Josephson tunnel barrier by
measuring the voltages of Fiske steps in long junctions made of
the same trilayer. The charging energy $E_{\rm C}=e^2/(2C)$
corresponding to the junction capacitance $C$ was about $0.91\cdot
10^{-23}\,$J, yielding the ratio of $E_{\rm J}/E_{\rm C}\approx
5.4\,$. Measurements have been performed at a temperature
$4.2\,$K,  yielding $k_{\rm B }T=5.8\cdot 10^{-23}\,$J. Thus, the
data presented here approximately fall into the parameter range of
$k_{\rm B }T\approx E_{\rm J}\gg E_{\rm C}\,$.

\begin{figure}
\includegraphics[width=3in]{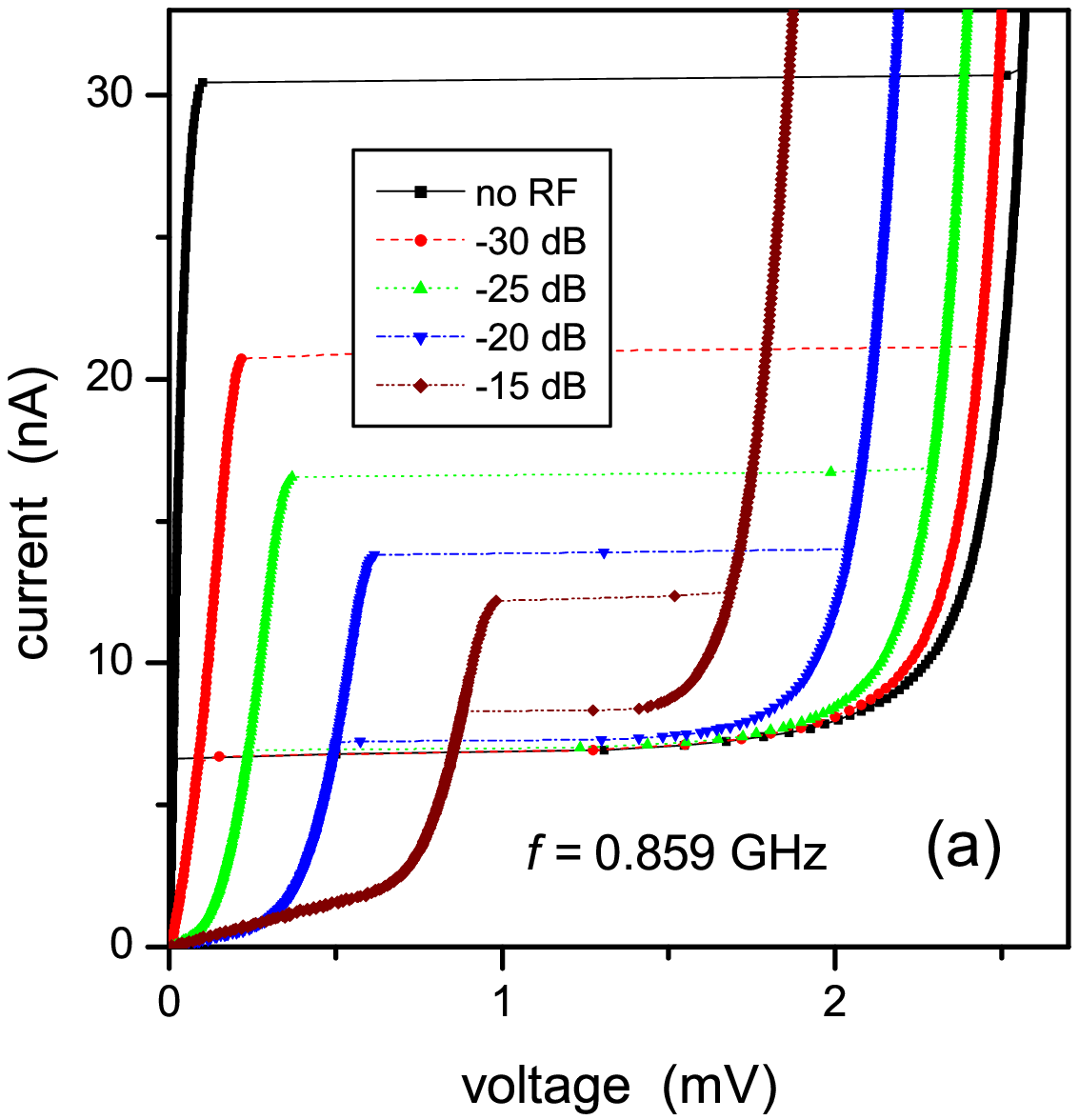}
\includegraphics[width=3in]{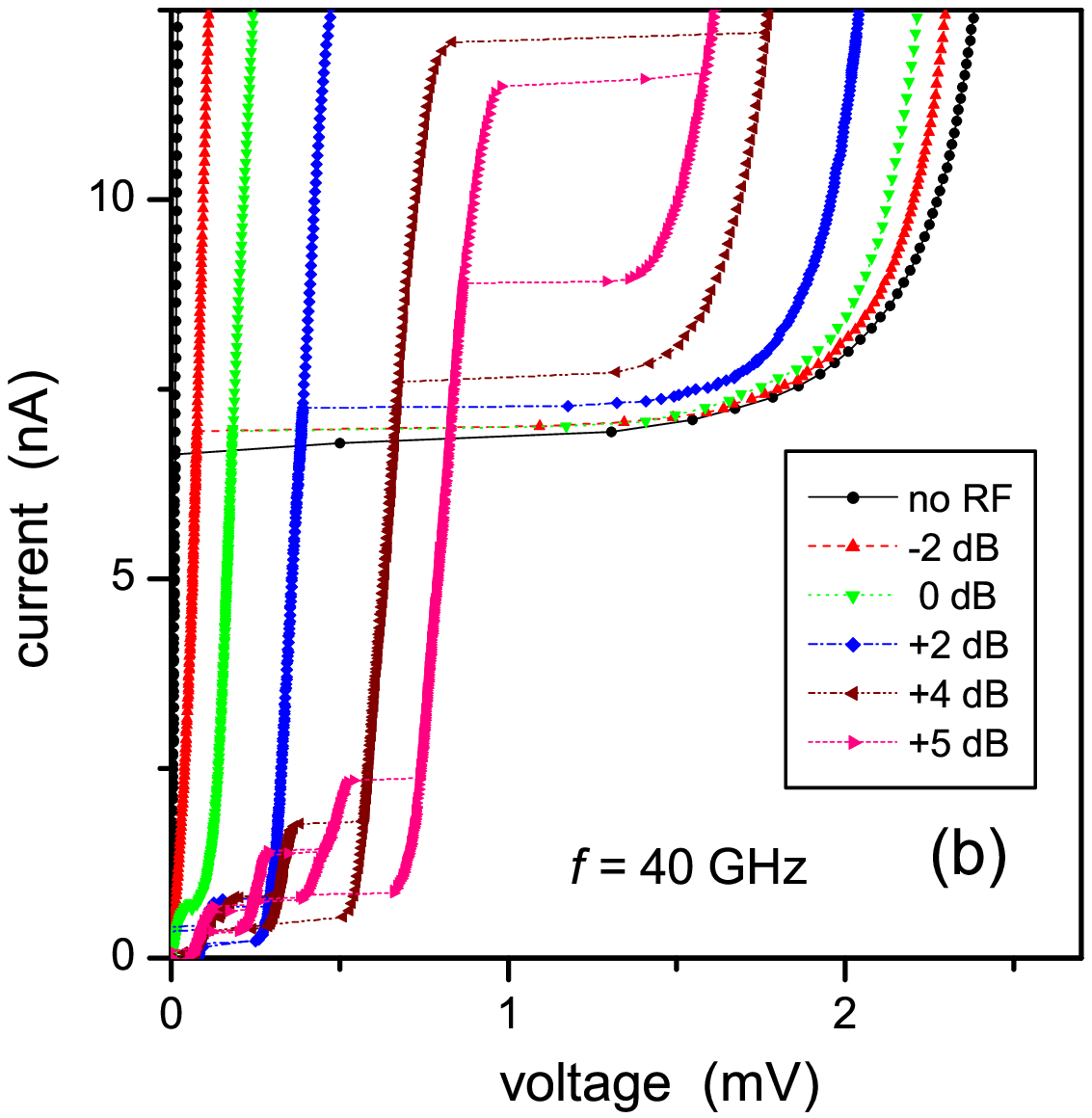}
\caption{Current-voltage characteristics of
Nb/AlO$_x$/Nb Josephson tunnel junction at $4.2\,$K. Different
curves show the effect of increasing microwave power of the
frequency $0.859\,$GHz (a) and $40\,$GHz (b). The microwave power
levels are referenced to the top of the cryostat.} \label{2IVsets}
\end{figure}

The measured $I$-$V$ curves of the junction under microwave
irradiation are shown in Fig.~\ref{2IVsets}. We present here data
for one polarity of the bias current, as the curves were found to
be symmetric with respect to the origin. Microwaves induce a
pronounced peak in the $I$-$V$ curve which shifts towards larger
voltages with increasing the microwave power. Simultaneously, we
observe the suppression of the superconducting energy gap branch,
which we interpret as the well-known photon-assisted quasiparticle
tunneling \cite{Tinkham}. For low frequency $f$ of microwaves, as
shown in Fig.~\ref{2IVsets}(a), both the peak voltage position and
its voltage range are much higher than the expected voltage
spacing between the Shapiro steps. As the microwave frequency $f$
increases, the low-voltage part of the peak splits into several
well-separated steps with finite slope, as seen in
Fig.~\ref{2IVsets}(b). The expected Shapiro step voltage spacing
here is $\Delta V=\Phi_0f\approx 83\,\mu$V, which matches the
observed magnitude of the voltage jumps between the low-current
features in the $I$-$V$ curves in Fig.~\ref{2IVsets}(b).

\begin{figure}
\includegraphics[width=3.4in]{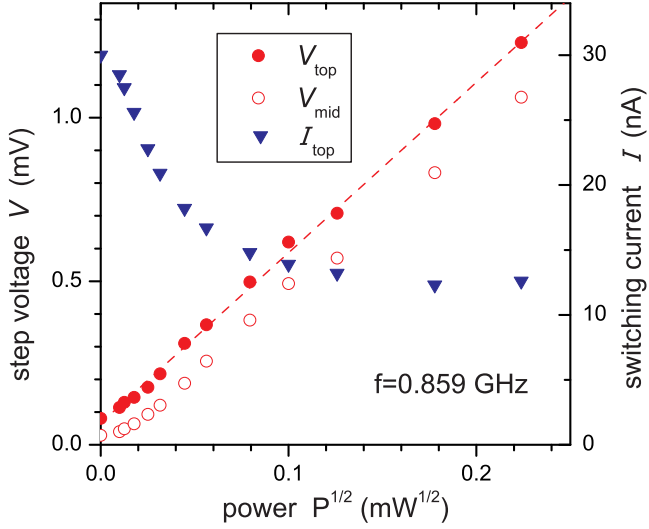}
\caption{Dependence of the voltages $V_{\rm top}$ (solid circles)
and $V_{\rm mid}$ (open circles) and the current $I_{\rm top}$ (triangles) on
microwave field amplitude $\propto\sqrt P$. The values of $V_{\rm
top}$  and $I_{\rm top}$  are measured
at the switching point on the top of the microwave-induced peak on
the $I$-$V$ characteristics. The voltage $V_{\rm mid}$
refers to the middle (half-height) point of the peak.
Dashed line is a linear fit to $V_{\rm top}(\sqrt P)$}.
\label{power}
\end{figure}

The voltage of the peak increases with the power $P$ of
microwaves. Figure \ref{power} shows the measured voltage $V_{\rm
top}$ at the switching point on the top of the peak as a function
of the microwave field amplitude $\propto\sqrt P$. One can see
that the voltage $V_{\rm top}$ rises proportionally to the
amplitude of the applied microwaves. The current amplitude of the
peak decreases with power, as summarized in Fig.~\ref{power}.

Next we turn to the analysis of the $I$-$V$ curve of a small
Josephson junction in the phase diffusion regime. A Josephson
junction is characterized by the Josephson phase $\varphi(t)$
satisfying the equation:
\begin{equation} \label{GenEq}
\ddot \varphi(t)+\alpha \dot \varphi(t)+\sin\varphi(t)= j+\eta
\sin\omega t +\xi(t)\;.
\end{equation}
In this equation, $j$ is an external dc bias, $\eta$ and $\omega$
are amplitude and frequency of an external microwave radiation,
$\xi(t)$ is a random function of time $t$ describing thermal
fluctuations. Here, we use dimensionless units, i.e. the time is
normalized to $\omega_p^{-1}$, the dc bias is normalized to the
critical current value $I_c$. The dissipative effects are
described by parameter $\alpha$ that, in general, can be frequency
dependent. The presence of hysteretic regime in the observed
$I$-$V$ curves allows to propose a simple model, assuming that the
low frequency damping $\alpha_0$ determining the quasiparticle
current is weak ($\alpha_0~\ll~1$) but the high frequency damping
$\alpha$ is large. As the Josephson coupling energy $E_{\rm J}$ is
comparable with the energy of thermal fluctuations $k_{\rm B }T$,
the incoherent superconducting current can be found from the
following analysis. In the absence of applied microwave radiation
the Josephson phase is written as
\begin{equation} \label{Josphase}
\varphi(t)=vt+\psi(t)+\varphi_1(t)~~,
\end{equation}
where $\psi(t)=\frac{1}{\alpha}\int_0^t \xi(t) dt$ determines the
{\it Josephson phase diffusion}. Here, the dc voltage drop across
the Josephson junction $V$ is normalized to
$V_p=\Phi_0\omega_p/(2\pi)=\frac{1}{2e}\sqrt{8E_{\rm J} E_{\rm C}}
=0.187 \,$mV as $v=V/V_p$. In this case, the term $\varphi_1(t)$
is expressed through the alternating part of the superconducting
current
\begin{equation} \label{Josphasecorr}
\varphi_1(t)~=~\frac{1}{\alpha}\int_0^t dt_1
\sin\left(vt_1+\psi(t_1)\right)~~.
\end{equation}
The dc component of the incoherent superconducting current is
found as
$$
I_s^0(v)=
$$
$$
\frac{I_c}{\alpha}\lim_{T\rightarrow\infty}\int_0^T
\frac{dt}{T}\cos\left(vt+\psi(t)\right) \int_0^t dt_1
\sin\left(vt_1+\psi(t_1)\right)=
$$
\begin{equation} \label{dcSuperCurrent}
\frac{I_c}{\alpha}\int_0^\infty dt  \rho(t) \sin(vt)~~,
\end{equation}
where $\rho(t)~=~<\cos(\psi(t)-\psi(0))>$ is the correlation
function of a random part of the Josephson phase. This correlation
function has been studied in detail for different models in
\cite{Ing-Grab-Eber-94,IngNaz} and, in the simplest case of
Gaussian random function $\xi(t)$, the $\rho(t)$ has a diffusive
form $\rho(t)~=~\exp(-\delta t)$. Substituting this expression
into (\ref{dcSuperCurrent}) we obtain \cite{Tinkham,IngNaz}
\begin{equation} \label{dcSCFin}
I_s^0~=~\frac{I_c}{\alpha}\frac{v}{v^2+\delta^2}~
=~\frac{I_c}{\alpha}\frac{VV_p}{V^2+(\delta V_p)^2}~~.
\end{equation}
This dependence is fitted to our $P=0$ data by using $\delta=0.53$
and $\alpha= 4.7\,$. It is shown in Fig. 3 by dashed line. As
expected, it is in a good agreement with the experimentally
measured $I$-$V$ curve in the absence of microwaves. We note that
the above value of $\delta$ translates into the high frequency
impedance seen by the junction $Z=\delta V_p\Phi_0/(2\pi k_{\rm B
} T)\approx 570\,\Omega$, which is not far from the expectation
that the bias leads impedance is typically of the order of
$100\,\Omega$.

In the presence of microwave radiation of a frequency
$f=\omega/(2\pi)$ the Josephson phase can be written as
\begin{equation} \label{JosphaseMicrwave}
\varphi(t)~=~vt+\psi(t)-\frac{\eta}{\alpha}\cos\omega
t+\varphi_1(t)~~.
\end{equation}
Applying the procedure that is similar to the above treatment of
the Josephson phase diffusion without microwave radiation, we
obtain the enhancement of superconducting current. This
enhancement is due to incoherent absorption (emission) of photons
by Josephson phase
\begin{equation} \label{SupercurrentMicroW}
I_s=\sum_{n=-\infty}^{\infty}J_n^{2}\left(\frac{\eta}{\alpha
\omega}\right)I_s^{0}(v-n\omega)~~,
\end{equation}
where $J_n(x)$ is the Bessel function of index $n$, and $n$ the
number of photons. By making use of the properties of Bessel
functions \cite{Stigan} and the limit of multiphoton absorption
($n\gg~1$) the superconducting current can be written in the
following form
\begin{equation} \label{SCpreobr}
I_s=\frac{1}{2\pi}\int_0^{\pi} du \left[I_s^0(v+\eta^\prime \cos
u)+ I_s^0(v-\eta^\prime \cos u)\right]~~,
\end{equation}
where $\eta^\prime=\eta/\alpha\propto\sqrt{P}$. By substituting
expression (\ref{dcSCFin}) into (\ref{SCpreobr}) we obtain that
the superconducting current displays a pseudo-resonant feature,
i.e. a peak with downward curvature. The voltage position $V_0$ of
this peak increases with the power of microwave radiation as
$V_{0}=V_p\eta^\prime\propto\sqrt{P}$. The equation for $I_s(V)$
can be presented in the analytical form:
\begin{equation} \label{SCanalitik}
I_s=I_c\frac{1}{2\alpha}
\left[f_+(v_-)f_+(v_+)-f_-(v_-)f_-(v_+)\right]~,
\end{equation}
where we introduced the notations $v_{\pm}=(V\pm V_0)/V_p$ and
$f_{\pm}(x)=\sqrt{( \sqrt{x^2+\delta^2} \pm x)/(x^2+\delta^2)} $ .

Thus, the maximum value of the superconducting current at $V_0\gg
V_p\delta$ weakly decreases with the microwave power $P$ as
\begin{equation} \label{MaximumCurr}
I_s^{\max}~=~I_c\frac{1}{2\alpha}\frac{1}{\sqrt{\eta^\prime
\delta}}~\propto~P^{-1/4}~~.
\end{equation}
The total $I$-$V$ curve is the sum of the superconducting  and
quasiparticle currents $I(V)=I_s(V)+\alpha_0 V$ (note here that
only the part of the curve with the positive derivative is
stable). However, the low frequency damping $\alpha_0$ is small
and we neglect the quasiparticle current $\alpha_0 V$ in the
following discussion.

\begin{figure}
\includegraphics[width=3.3in]{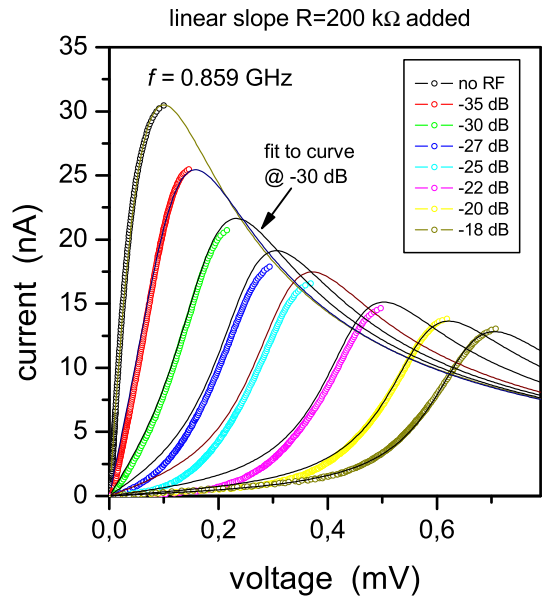}
\caption{Comparison between our theoretical model and experiment
for the frequency $f=0.859\,$GHz. Dots are experimentally measured
data; the dashed line corresponds to Eq. (\ref{dcSCFin}) and
solid lines correspond to the theoretical prediction
according to Eq.~(\ref{SCanalitik}). } \label{fit}
\end{figure}

Figure~\ref{fit} presents a comparison of the experimental data
with theory for the frequency $f=0.859\,$GHz. Solid lines
correspond to the theoretical prediction according to
Eq.~(\ref{SCanalitik}). Our analysis is in excellent agreement
with the observations. The theoretical curves in the presence of
microwaves are obtained by fitting a \emph{single} experimental
data set for the microwave power level of $P_0=-30$ dB and by
calculating \emph{all} other curves according to their power
levels relative to $P_0$. Note that we used for this procedure the
parameters $\delta$ and $\alpha$ obtained from fitting the $P=0$
data to Eq.~(\ref{dcSCFin}) as mentioned above.

As the microwave frequency $f$ increases, the smaller values of
$n$ contribute to the expression (\ref{SupercurrentMicroW}) and
the discrete Shapiro steps having the \emph{finite slope} appear
on the $I$-$V$ curve, see Fig.~\ref{2IVsets}(b). Our analysis
based on Eqs.~(\ref{SCpreobr}) and (\ref{dcSuperCurrent}) can be
extended to describe the Josephson phase diffusion in more complex
cases. It can be applied to Josephson junctions embedded in a
frequency dependent environment, e.g. in the presence of
resonances in an external circuit \cite{Vion-96}, or to a weak
Josephson phase diffusion with voltage-dependent parameter
$\delta$. As the parameter $\delta$ is small, i. e.
$\delta~\ll~\omega$ , the latter regime is of particular interest
because the analysis of Eq.~(\ref{SupercurrentMicroW}) shows that
at some ac power {\em zero-crossing steps} can appear in the
$I$-$V$ curves \cite{KFUunp}. Moreover, we have observed such
incoherent zero-crossing steps at 4.2 K for Josephson junctions of
an intermediate size with $E_J~\geq~k_B T $ \cite{KFUunp}. Similar
incoherent zero-crossing steps have been also seen in have been
also seen in intrinsic junctions based on high-temperature
superconductors.

In summary, we have measured the effect of microwaves on
mesoscopic Josephson junctions with characteristic energies
$E_{\rm C}\ll E_{\rm J}\approx k_{\rm B }T \,$. In contrast to the
behavior of larger Josephson junctions with $E_{\rm J}\gg k_{\rm B
}T \,$, which display well-known discrete Shapiro steps due to
locking of their Josephson oscillations to microwave frequency,
for small junctions we find a smooth and \emph{incoherent
enhancement of Josephson phase diffusion by microwaves}. This
enhancement is manifested by a pronounced current peak at the
voltage $V_{\rm top}\propto\sqrt P$. We have proposed here a
theoretical model for the microwave-stimulated phase diffusion
which showed excellent agreement with the measured data.

We would like to thank A. Abdumalikov, D. Averin, J. Clarke , E.
Ilichev, S. Shitov, and H.B. Wang for useful discussions and
acknowledge the partial financial support by the Deutsche
Forschungsgemeinschaft (DFG).

\end{document}